\documentclass[%
 reprint,
 amsmath,
 amssymb,
 aps,
]{revtex4-1}

\usepackage{mathtools}
\usepackage[caption = false]{subfig}
\usepackage{graphicx}% Include figure files
\graphicspath{{images/}}

\usepackage{dcolumn}% Align table columns on decimal point
\usepackage{bm}% bold math

\usepackage{hyperref} % add hypertext capabilities
\begin{document}

%\preprint{APS/123-QED}

\title{From periodically driven double wells to volcano potentials: Quantum dynamics}

\author{Soumyabrata Paul} 
\email{soumyabrata.paul93@gmail.com}
\author{Rohit Chawla}
\author{Mandira Das}
\author{Jayanta K. Bhattacharjee}
\affiliation{Department of Theoretical Physics, Indian Association for the Cultivation of Science, Jadavpur, Kolkata 700032, India}

\date{\today}

%%%%%%%%%%%%%%%% ABSTRACT %%%%%%%%%%%%%%%%%%%%%%%%
\begin{abstract}

We consider the dynamics of a particle confined in a double well potential which is subjected to a periodic drive. In the case of deep and well separated wells, we find that by adjusting the parameters of the drive we can generate, to a very good approximation, a volcano potential. The quantum dynamics in this volcano potential is studied by a variation of what can be called a generalized Ehrenfest's theorem. We find that the coupling of the mean position and the width of the wave packet in this dynamics causes the particle to escape from the central well in accordance with the fact that the volcano potential only supports resonance states. 

%\begin{description}	
%use the optional argument of the \verb+\item+ command to give the category of %each item. 
%\end{description}

\end{abstract}

\pacs{Valid PACS appear here}% PACS, the Physics and Astronomy
\maketitle

%\tableofcontents

%%%%%%%%%%%%% INTRODUCTION %%%%%%%%%%%%%%%%%
\section{Introduction} \label{introduction}

\begin{figure}[h]
	\includegraphics[width = \linewidth]{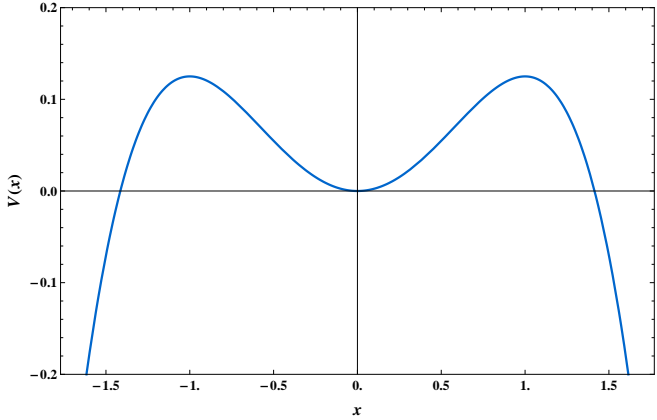}
	\caption{Volcano Potential. Numerical plot of $V_s(x)$ (see lines below Eq. \eqref{eq:9}) with $\omega^2 = 1$, $\lambda = 0.1$ and $\frac{\epsilon^2 \omega^2}{\Omega^2} = 3$.}
	\label{fig:fig1}
\end{figure}

A Volcano potential stands for a dip (value of the potential will be chosen as zero at the minimum) at the center (taken to be the origin of our coordinate system) with its value, in one dimension, approaching negative infinity or zero as $x$ approaches $\pm ~ \infty$, i.e. on the either side of the origin. The dip at the center with finite maxima on both sides gives the appearance of a volcano and we will restrict ourselves to the symmetric case shown in Fig. \ref{fig:fig1} with the potential falling off to negative infinity.

Volcano potentials \cite{ref1} appear in both condensed matter physics \cite{ref2} and high energy physics \cite{ref3, ref4, ref5}. In a very different context, periodically driven systems \cite{ref6}  have been extensively studied over the last several years and have given rise to a large number of interesting results both in classical \cite{ref7, ref8, ref9} and quantum regimes \cite{ref10, ref11}. In this work, we consider a double well potential driven by a very high frequency vibrations. A classical analysis of the problem will show that under certain conditions involving the amplitude and frequency of the periodic drive, it is possible to generate a volcanic potential of from a double well potential. In the quantum case, the corresponding dynamics should in principle be the dynamics of an initial wave packet and there have been several attempts of late \cite{ref12, ref13, ref14} to study quantum dynamics in the presence of barriers and quantum resonances. Here we adopt a different approach of looking at the dynamics following a generalization of Erhenfest's theorem to account for the quantum fluctuations \cite{ref15}. Consequently we find that what would be a bound orbit inside the central dip of the potential becomes an escaping orbit due to quantum fluctuations. This is consistent with the fact that the quantum states in the volcanic potential are the type considered here are necessarily resonant states. 

We begin with a double well potential given by,

\begin{equation} \label{eq:1}
V(x) = -\frac{m \omega^2 x^2}2 + \frac{\lambda m x^4}4
\end{equation}   

We will be primarily interested in small value of $\lambda$ for which the potential $V(x)$ will have a very pronounced double well structure as shown in Fig. \ref{fig:fig1}. We modulate this potential by a periodic time dependent function $f(t)$, so that the Hamiltonian of a particle in this potential is given by,

\begin{equation} \label{eq:2}
\mathcal{H} = \frac{p^2}{2m} + \left( 1 + \epsilon f(t) \right) V(x)
\end{equation} 

where $f(t)$ is a periodic function with time period $T = \frac{2 \pi}{\Omega}$. The frequency $\Omega$ will be taken to be high, i.e. $\Omega \gg \omega$ and $\frac{\epsilon \omega^2}{\Omega^2} \ll 1$. Our strategy to study the quantum dynamics will be to write the equations of motion for the dynamics of the different moments of the position operator and then carry out the separation of the \textit{fast} and \textit{slow} variables and directly determine the dynamics of the slow variables by averaging out the fast variables \cite{ref7}. This is a somewhat different viewpoint from the effective Hamiltonian strategy adopted by Rahav \textit{et al.} \cite{ref16} and Gilary \textit{et al.} \cite{ref17}. Our approach has the advantage of showing very clearly the difference between the classical and the quantum situations and in some sense sets up a different perspective for examining quantum dynamics. In Section \ref{classical}, we will examine the dynamics from the classical point of view and arrive at results which are in agreement with the corresponding results obtained by Blackburn et al. \cite{ref8} in a slightly different context. In Section \ref{quantum}, we present the quantum dynamics and the corresponding numerical results in Section \ref{numerics}. We finish with some comments in Section \ref{conclusion}.

%%%%%%%%%%%%%% CLASSICAL DYNAMICS %%%%%%%%%%%%%%%%%%%%%
\section{Classical Dynamics} \label{classical}

Our starting point is the classical equation of motion corresponding to the Hamiltonian of Eq. \eqref{eq:1} which reads, with $f(t)$ chosen as $\epsilon \cos \Omega t$, as follows

\begin{equation} \label{eq:3}
\ddot x = \omega^2 (1 + \epsilon \cos \Omega t) x - \lambda (1 + \epsilon \cos \Omega t) x^3
\end{equation}

We split $x(t)$ as

\begin{equation} \label{eq:4}
x(t) = x_s(t) + x_f(t)
\end{equation}

where $x_s(t)$ involves all the dynamics with the time scale $\frac{2 \pi}{\omega}$ and $x_f(t)$ contains the dynamics with the fast time scale $\frac{2 \pi}{\Omega}$. Our interest is in finding the dynamics of $x_s(t)$ by averaging out the fast time scales. To this end we substitute for $x(t)$ in Eq. \eqref{eq:3} from Eq. \eqref{eq:4} and obtain

\begin{align}
\begin{split} \label{eq:5}
\ddot x_s + \ddot x_f = {} & \omega^2 x_s - \lambda x_s^3 + \epsilon \omega^2 \cos \Omega t x_f \\
& - 3 \epsilon \lambda \cos \Omega t x_s^2 x_f - \epsilon \lambda \cos \Omega t x_f^3 \\
& - 3 \lambda x_f^2 x_s + \bigg[ \omega^2 x_f  - \lambda x_f^3 \\
& + \epsilon \omega^2 \cos \Omega t x_s - \epsilon \lambda \cos \Omega t x_s^3 \\
& - 3 \lambda x_s^2 x_f - 3 \epsilon \lambda \cos \Omega t x_f^2 x_s \bigg]
\end{split}
\end{align}

We expect $x_f(t) \approx \cos \Omega t$ and based on that, note that the terms in the square bracket are all the fast terms, whereas all the other terms vary on the slow time scale. Accordingly

\begin{subequations} \label{eq:6}
	\begin{align}
	& \ddot x_s = \omega^2 x_s - \lambda x_s^3 - 3 \lambda x_f^2 x_s + \epsilon \omega^2 \cos \Omega t x_f  \nonumber \\
	& \qquad - 3 \epsilon \lambda \cos \Omega t x_s^2 x_f - \epsilon \lambda \cos \Omega t x_f^3  \label{eq:6a} \\		
	& \ddot x_f = \omega^2 x_f + \epsilon \omega^2 \cos \Omega t x_s - \lambda x_f^3 - \epsilon \lambda \cos \Omega t x_s^3 \nonumber \\
	& \qquad - 3 \epsilon \lambda x_f^2 x_s \cos \Omega t - 3 \lambda x_s^2 x_f \label{eq:6b}		
	\end{align}
\end{subequations}

We will solve $x_f(t)$ to the lowest order in $\epsilon$ so that Eq. \eqref{eq:6b} can be linearized in $x_f(t)$ and we get

\begin{equation} \label{eq:7}
\ddot x_f - \omega^2 x_f + 3 \lambda x_s^2 x_f = \epsilon \omega^2 \cos \Omega t x_s - \epsilon \lambda \cos \Omega t x_s^3
\end{equation}

leading to ($\Omega \gg \omega$)

\begin{equation} \label{eq:8}
x_f = -\frac{\epsilon}{\Omega^2}(\omega^2 x_s - \lambda x_s^3) \cos \Omega t
\end{equation}

Inserting the above $x_f(t)$ in Eq. \eqref{eq:6a} and averaging over time, we get to leading order in $\Omega^{-1}$

\begin{equation} \label{eq:9}
\ddot x_s = \omega^2 \left(1 - \frac{\epsilon^2 \omega^2}{2 \Omega^2} \right) x_s - \lambda \left(1 - \frac{2 \epsilon^2 \omega^2}{\Omega^2} \right) x_s^3
\end{equation}

If $\frac{\epsilon^2 \omega^2}{2 \Omega^2} > 1$, then the effective potential for the slow dynamics is $V_s(x) = \alpha \frac{x_s^2} 2 - \beta \frac{x_s^4} 4$, where $\alpha \equiv \omega^2 \left(\frac{\epsilon^2 \omega^2}{2 \Omega^2} - 1 \right)$ and $\beta \equiv \lambda \left(\frac{2 \epsilon^2 \omega^2}{\Omega^2} - 1 \right)$ are both positive numbers. Thus in the classical case the time dependent double well potential becomes a volcano potential for the slow variable for some parameter values.

%%%%%%%%%%%%%%%%% QUANTUM DYNAMICS %%%%%%%%%%%%%%%%%%5
\section{Quantum Dynamics} \label{quantum}

We now treat the Hamiltonian in Eq. \eqref{eq:1} quantum mechanically and ask what would be the quantum dynamics of a particle in this high frequency limit. The quantum dynamics involves starting out with an initial wave packet $\Psi(x, 0)$ and finding what the wave packet would be at time $t$. In general this is very difficult to accomplish even in the simple situations. We will take the point of view that the dynamics will be revealed by the mean position of the particle and the mean spread in its position. If one writes down the relevant time dependent equations for arbitrary moments(a generalization of Erhenfest's Theorem) there will be an infinite hierarchy of coupled equations which we can truncated by an appropriate closure scheme. In a whole variety of physical problems, this situation of a coupled set of dynamical equations for infinite number of moments exists. Some of the well known ones are in turbulence \cite{ref18, ref19} and kinetics of phase separations \cite{ref20, ref21}. Truncating the hierarchy \cite{ref22, ref23} by some closure assumption is most often the only available path. We will analyze the coupled dynamics to see how the particle will escape from inside the classically stable region of the volcano due to quantum fluctuations. We will begin with Erhenfest's equations for $\left \langle x \right \rangle$ and $\left \langle p \right \rangle$

\begin{subequations} \label{eq:10}
	\begin{align}
	& \frac{d}{dt} \left \langle x \right \rangle = \frac{\left \langle p \right \rangle}{m} \label{eq:10a} \\
	& \frac{d}{dt} \left \langle p \right \rangle = m \omega^2 \left( 1 + f \right) \left \langle x \right \rangle - \lambda m \left( 1 + f \right) \left \langle x^3 \right \rangle \label{eq:10b}	
	\end{align}
\end{subequations}

leading to

\begin{equation} \label{eq:11}
\begin{split} %requires amsmath package
\frac{d^2}{dt^2} \left \langle x \right \rangle 
&= \omega^2 \left(1 + f \right) \left \langle x \right \rangle - \lambda \left( 1 + f \right) \left \langle x^3 \right \rangle \\ 
&= \omega^2 \left( 1 + f \right) \left \langle x \right \rangle - \lambda \left( 1 + f \right) \left \langle x \right \rangle^3 \\ \MoveEqLeft[-1] - \lambda \left( 1 + f \right) \left[ \left \langle x^3 \right \rangle - \left \langle x \right \rangle^3 \right]
\end{split}	
\end{equation}

If we ignore the terms in the square brackets in the above equations, then we have the classical equation of motion for the dynamics of $\left \langle x \right \rangle$. This, as expected, is identical to the classical dynamics given by Eq. \eqref{eq:3}. What should be noted is that this is not the whole story and the term in the square bracket does not vanish. The fact that here the quantum fluctuations  $\left \langle x^3 \right \rangle -  \left \langle x \right \rangle^3$ is nonzero and has its own dynamics is what makes the quantum dynamics different. We need to know what the dynamics of $\left \langle x^3 \right \rangle -  \left \langle x \right \rangle^3$ is and needless to say thus we bring in the higher moments. Hence it is good to understand the features of $\left \langle x^3 \right \rangle -  \left \langle x \right \rangle^3$ before proceeding any further. We note

\begin{equation} \label{eq:12}
\begin{split} %requires amsmath package		
\left \langle x^3 \right \rangle -  \left \langle x \right \rangle^3 		
&= \left \langle \left[ x - \left \langle x \right \rangle + \left \langle x \right \rangle \right]^3 \right \rangle - \left \langle x \right \rangle^3 \\
&= \left \langle \left( x - \left \langle x \right \rangle \right)^3 \right \rangle + 3 \left \langle \left( x - \left \langle x \right \rangle \right)^2 \right \rangle \left \langle x \right \rangle
\end{split}
\end{equation}

The first term on the right hand side is the skewness. We will assume for simplicity that the skewness remains zero if we start with a wave function symmetric about its center so that the skewness is initially zero(this approximation will certainly breakdown at boundaries of the classical region). Within this approximation,

\begin{equation} \label{eq:13}
\left \langle x^3 \right \rangle - \left \langle x \right \rangle^3 = 3 W \left \langle x \right \rangle
\end{equation}

where $W \equiv \left \langle x^2 \right \rangle - \left \langle x \right \rangle^2$ is the mean square width of the wave packet. The dynamics of $W$ is found from Ehrenfest's Theorem as

\begin{align} \label{eq:14}
\frac{d^2}{dt^2} W &= \frac{2}{m^2} \left[ \left \langle p^2 \right \rangle - \left \langle p \right \rangle^2 \right] + 2 \omega^2 (1 + f) W \nonumber \\
& \quad - 2 \lambda (1 + f) \left[ \left \langle x^4 \right \rangle - \left \langle x^3 \right \rangle \left \langle x \right \rangle \right]
\end{align}

With vanishing skewness, $\left \langle x^4 \right \rangle - \left \langle x^3 \right \rangle \left \langle x \right \rangle = K + 3 W \left \langle x \right \rangle^2$, where $K \equiv \left \langle \left( x - \left \langle x \right \rangle \right)^4 \right \rangle$ is the kurtosis of the distribution. If we now write down the dynamics it will couple to the sixth order cumulant $\left \langle \left( x - \langle x \rangle \right)^6 \right \rangle$ and higher. For closure at this stage, we make the Gaussian approximation $K = 3 W^2$. Applying Ehrenfest's Theorem to $\left \langle \left( \Delta p \right)^2 \right \rangle \equiv \left \langle p^2 \right \rangle - \left \langle p \right \rangle^2$, we get

\begin{align} \label{eq:15}
\frac{d}{dt} \left\langle (\Delta p)^2 \right\rangle &= m^2 \omega^2 \left( 1 + f \right) \frac{d W}{dt} \nonumber \\
& \quad - \frac{\lambda}{2} m^2 \left( 1 + f \right) \left[ 3 \frac{dW^2}{dt} + 6 \langle x \rangle^2 \frac{dW}{dt} \right]
\end{align}

and using this in Eq. \eqref{eq:14}, we finally get

\begin{align} \label{eq:16}
\frac{d^3}{dt^3} W &= 4 \omega^2 (1 + f) \frac{dW}{dt} + 2 \omega^2 \dot f W - 9 \lambda (1 + f) \frac{d}{dt} W^2 \nonumber \\ 
& \quad  - 12 \lambda (1 + f) \left \langle x \right \rangle^2 \frac{dW}{dt} - 6 \lambda \dot f \left[ W^2 + W \left \langle x \right \rangle^2 \right] \nonumber \\
& \quad  - 6 \lambda (1 + f) W \frac{d}{dt} \left \langle x \right \rangle^2
\end{align}

We write the dynamics of $\langle x \rangle$ as (see Eq. \eqref{eq:11}) 

\begin{align} \label{eq:17}
\frac{d^2}{dt^2} \langle x \rangle &= \omega^2 (1 + f) \langle x \rangle - \lambda (1 + f) \left \langle x \right \rangle^3 \nonumber \\
& \quad - 3 \lambda (1 + f) W \langle x \rangle	
\end{align}

The equations of motion are now in a closed set represented by Eq. \eqref{eq:16} and Eq. \eqref{eq:17}. At this point we will set $f(t) = \epsilon \cos \Omega t$ as we did in Section \ref{classical}. With $W = 0$, we have no quantum effect at all. The effect of quantum fluctuations is provided by $W$. To make progress we split $\langle x \rangle$ and $W$ into slow and fast parts by writing

\begin{subequations} \label{eq:18}
	\begin{align}
	\langle x \rangle &= \langle x\rangle_s + \langle x \rangle_f \\
	\langle W \rangle &= \langle W\rangle_s + \langle W \rangle_f 
	\end{align}
\end{subequations}

This leads to

\begin{align} \label{eq:19}
\ddot {\langle x \rangle}_s + \ddot {\langle x \rangle}_f &= \omega^2 \left( \langle x \rangle_s + \langle x \rangle_f \right) + \epsilon \omega^2 \cos \Omega t ( \langle x \rangle_s \nonumber \\ 
& \quad + \langle x \rangle_f ) - \lambda \left(1 + \epsilon \cos \Omega t \right) \big( \langle x \rangle_s^3 \nonumber \\
& \quad + 3 \langle x \rangle_s^2 \langle x \rangle_f \big) - 3 \lambda \left( 1 + \epsilon \cos \Omega t \right) \nonumber \\
& \qquad \big( \langle x \rangle_s W_s + \langle x \rangle_f W_s + \langle x \rangle_s W_f \big)
\end{align}

and

\begin{align} \label{eq:20}
\dddot W_s + \dddot W_f &= 4 \omega^2 \left(1 + \epsilon \cos \Omega t \right) \left( \dot W_s + \dot W_f \right) \nonumber \\
& \quad - 2 \epsilon \omega^2 \Omega \sin \Omega t \left( W_s + W_f \right) \nonumber \\
& \quad - 9 \lambda \left( 1 + \epsilon \cos \Omega t \right) \frac{d}{dt} \left( W_s + W_f \right)^2 \nonumber \\
& \quad - 12 \lambda \left( 1 + \epsilon \cos \Omega t \right) \left( \langle x \rangle_s + \langle x \rangle_f \right)^2 \nonumber \\ 
& \qquad \frac{d}{dt} \left( W_s + W_f \right) + 6 \epsilon \lambda \Omega \sin \Omega t \nonumber \\
& \qquad \bigg[ \left( W_s + W_f \right)^2 + \left( W_s + W_f \right) \nonumber \\
& \qquad \left( \langle x \rangle_s + \langle x \rangle_f \right)^2 \bigg] - 6 \lambda \left( 1 + \epsilon \cos \Omega t \right) \nonumber \\
& \qquad \left( W_s + W_f \right) \frac{d}{dt} \left( \left \langle x \right \rangle_s + \left \langle x \right \rangle_f \right)^2
\end{align}

%Repeating exactly the procedure followed in arriving at \eqref{eq:9} starting from %\eqref{eq:3}, we obtain after long but straightforward algebra(working to $\mathcal{O}( %\lambda)$)

We need to go through the procedure we adopted to arrive at Eq. \eqref{eq:9} starting from Eq. \eqref{eq:3}. Accordingly, we separate the dynamics of Eq. \eqref{eq:19} and Eq. \eqref{eq:20} into fast and slow components. Solving for $x_f$ and $W_f$ we get

\begin{equation} \label{eq:21}
\langle x \rangle_f = - \frac{\epsilon}{\Omega^2} \left[ \omega^2 \langle x \rangle_s - \lambda \left( \langle x \rangle_s^3 + 3 \langle x \rangle_s W_s \right) \right] \cos \Omega t
\end{equation}

and 

\begin{align} \label{eq:22}
W_f &= - \frac{2 \epsilon}{\Omega^2} \left[ \omega^2 W_s - 3 \lambda \left( W_s \langle x \rangle_s^2 + W_s^2 \right) \right] \cos \Omega t \nonumber \\
& \quad - \frac{2 \epsilon}{\Omega^3} \bigg[ 2 \omega^2 \dot W_s - 3 \lambda \bigg( \frac{3}{2} \dot W_s^2 + W_s \langle \dot x \rangle_s^2 \nonumber \\
& \qquad + 2 \dot W_s \langle x \rangle_s^2 \bigg) \bigg] \sin \Omega t
\end{align}

Using the above in the dynamics of $\langle x \rangle_s$ and $W_s$, we finally arrive at the equations of motion(working to $\mathcal{O(\lambda)}$)

\begin{align} \label{eq:23}
\ddot{\langle x \rangle}_s &= \omega^2 \left( 1 - \frac{\epsilon^2 \omega^2}{2 \Omega^2} \right) \langle x \rangle_s - \lambda \left( 1 - \frac{2 \epsilon^2 \omega^2}{\Omega^2} \right) \langle x \rangle_s^3 \nonumber \\
& \quad - 3 \lambda \left( 1 - \frac{2 \epsilon^2 \omega^2}{\Omega^2} \right) \langle x \rangle_s W_s	 
\end{align}

and

\begin{align} \label{eq:24}
\dddot W_s &= 4 \omega^2 \left( 1 - \frac{2 \epsilon^2 \omega^2}{\Omega^2} \right) \dot W_s - 9 \lambda \left( 1 - \frac{5 \epsilon^2 \omega^2}{\Omega^2} \right) \dot W_s^2 \nonumber \\
& \quad - 12 \lambda \left( 1 - \frac{9 \epsilon^2 \omega^2}{\Omega^2} \right) \langle x \rangle_s^2 \dot W_s \nonumber \\
& \quad - 6 \lambda \left( 1 - \frac{5 \epsilon^2 \omega^2}{\Omega^2} \right) \langle \dot x \rangle_s^2 W_s
\end{align}

Clearly, the effective potential for $\langle x \rangle_s$(if we ignore the coupling to $W_s$) is a volcano potential for $\frac{\epsilon^2 \omega^2}{2 \Omega^2} > 1$. It has the form $V_s(\langle x \rangle_s) = \alpha \frac{\langle x \rangle_s^2}{2} - \beta \frac{\langle x \rangle_s^4}{4}$ as found in the classical case (see lines below Eq. \eqref{eq:9}). However, this is quantum dynamics and there is the additional term coming from the coupling to the width in Eq. \eqref{eq:23}. We note that the effective potential has peaks at $\langle x \rangle_s = \pm \sqrt \frac{\alpha}{\beta}$ and the heights of the peaks are equal having the value $\frac{\alpha^2}{4 \beta}$. Since $\beta = - \lambda \left( 1 - \frac{2 \epsilon^2 \omega^2}{\Omega^2} \right)$, the peaks are very far away from the center $\left( \langle x \rangle_s \right)$ and have very large values for $\lambda \rightarrow 0$. The classical dynamics will be confined in between $- \sqrt \frac{\alpha}{\beta}$ and $+ \sqrt \frac{\alpha}{\beta}$, so long as the total energy(conserved quantity) is less than $\frac{\alpha^2}{4 \beta}$. This implies for zero initial momentum, the dynamics will always be confined if the initial displacement($x_0$) is less than $\sqrt \frac{\alpha}{\beta}$. However, in this quantum case an initial wave packet with zero average momentum and centered well within the volcano will be able to escape from the well after a sufficiently long time. Our results will have some inaccuracy because of our neglect of the skewness which has been taken to be zero at all time. In reality as the wave packet moves towards $x = \pm \sqrt \frac{\alpha}{\beta}$ it will cease to be symmetric; and the consequent skewness will help a wave packet(centered at points very close to $x = 0$ at $t = 0$) to cross the barrier. This is in accord with the fact that there are no bound states in volcano potential of the type shown in Fig. \ref{fig:fig1}.

%%%%%%%%%%%% ESCAPE FROM THE WELL %%%%%%%%%%%%%
\section{Escape from the well} \label{numerics}

\begin{figure}[h]
	\includegraphics[width = \linewidth]{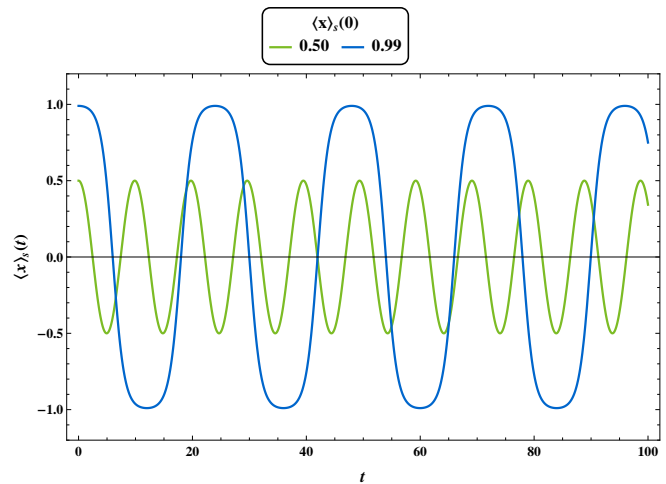}
	\caption{Numerical solution of the uncoupled dynamics of $\langle x \rangle_s$ with $\omega^2 = 1$, $\lambda = 0.1$, $\frac{\epsilon^2 \omega^2}{\Omega^2} = 3$ and $\langle x \rangle_s(0) = 0.5, 0.99$, $\langle \dot x \rangle_s(0) = 0$. The turning point of $V_s(\langle x \rangle_s)$ for these set of parameters are $\pm 1$ with height $0.125$.}
	\label{fig:fig2}
\end{figure}

In this section we study numerically the coupled system shown in Eq. \eqref{eq:23} and Eq. \eqref{eq:24}. To begin with, we note that if we ignore the coupling between $\langle x \rangle_s$ and $W_s$ completely, then $\langle \ddot x \rangle_s = - \frac{d V_s(\langle x \rangle_s)}{d \langle x \rangle_s}$. If we start with zero initial momentum and the position displaced from $\langle x \rangle_s = 0$, then so long as $\langle x \rangle_s(t = 0) < \sqrt{\frac \alpha \beta}$, the dynamics of $\langle x \rangle_s$ is oscillatory with increasing time period as the initial conditions approaches the turning point. Our numerical time series shown in Fig. \ref{fig:fig2} exhibits clearly the oscillation and the very long time period when the the initial condition is close to the turning point.  

Now we include the coupling between $\langle x \rangle_s$ and $W_s$ in Eq. \eqref{eq:23} but ignore the coupling in Eq. \eqref{eq:24}. This is motivated by the fact that the primary effect of the quantum fluctuations will be felt in the dynamics of the mean position and hence it is most important to keep the coupling to the dynamics of $\langle x \rangle_s$. We anticipate the coupling in the dynamics of $W_s$ to have a much smaller effect. We start with zero initial momentum and position displaced from $\langle x \rangle_s = 0$. We see that the dynamics of $\langle x \rangle_s$ is oscillatory with increasing time period as the initial position is moved more and more away from the center. However due to the coupling of $\langle x \rangle_s$ and $W_s$ in Eq. \eqref{eq:23} we note that the particle can exceed $\sqrt{\frac \alpha \beta}$ as opposed to the classical system. We see the effect of quantum fluctuations at play here which assists the particle to escape the well.

%%%%%%%%%%%%%%%%%% NUMERICAL PLOT OF PARTICALLY COUPLED SYSTEM %%%%%%%%%%%
\begin{figure}[h!]
	\includegraphics[width = \linewidth]{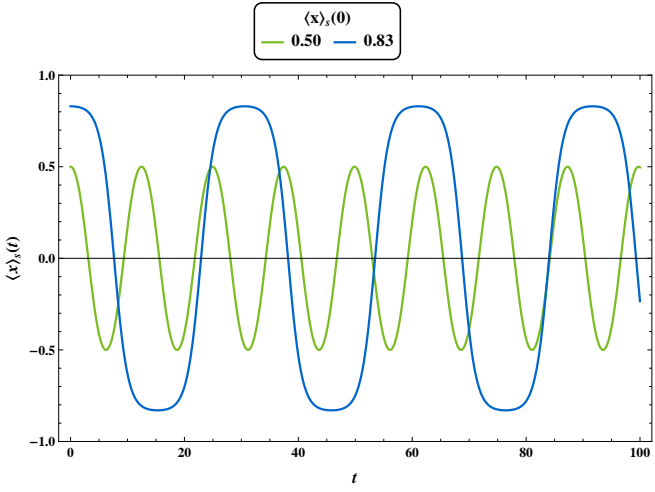}
	\caption{Numerical solution of the partially coupled dynamics of $\langle x \rangle_s$ (ignoring $\langle x \rangle_s$ coupling in the dynamics of $W_s$) with $\omega^2 = 1$, $\lambda = 0.1$, $\frac{\epsilon^2 \omega^2}{\Omega^2} = 3$ and $\langle x \rangle_s(0) = 0.5, 0.83$, $\langle \dot x \rangle_s(0) = 0$; with $W_s(0)= 0.1, \dot W_s(0) = 0$ and $\ddot W_s(0) = 0.01$.}
	\label{fig:fig3}
\end{figure}

\begin{figure}[b!]
	\begin{tabular}{cc}
		\subfloat[$\lambda = 0.5$]{\includegraphics[height = .3 \linewidth, width = .4 \linewidth]{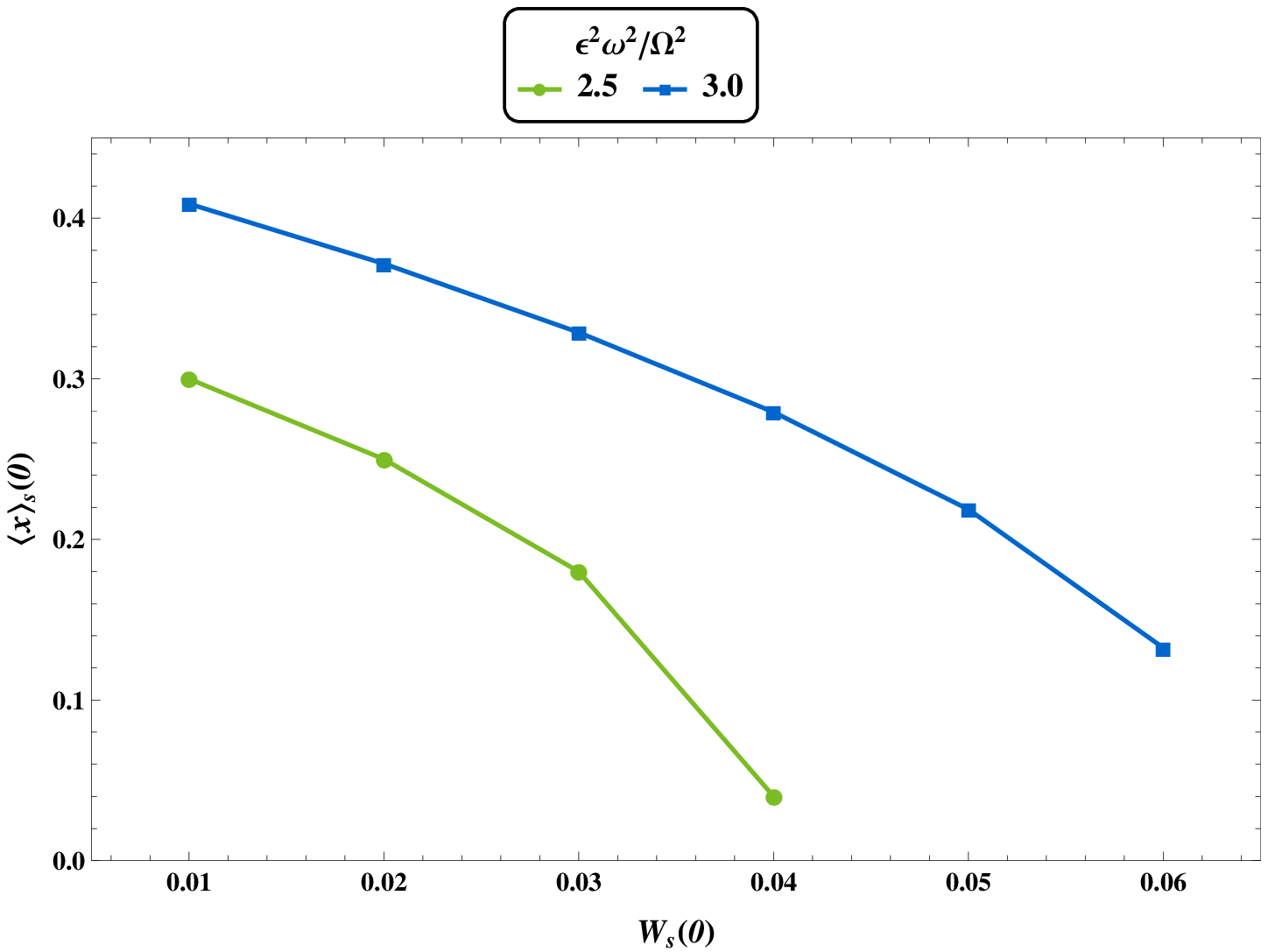} \label{fig:fig4a}} &		\subfloat[$\lambda = 0.1$]{\includegraphics[height = .3 \linewidth, width = .4 \linewidth]{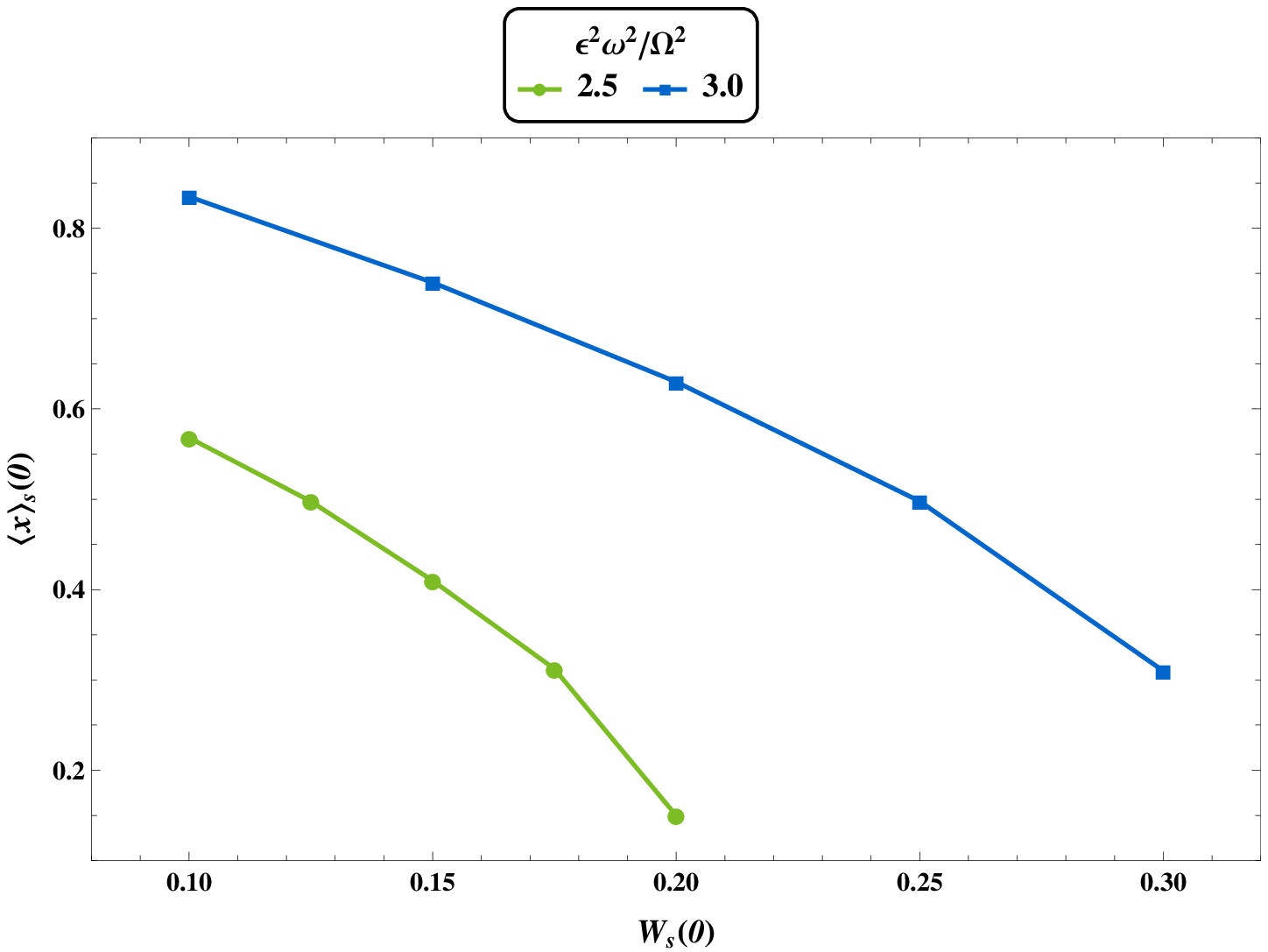} \label{fig:fig4b}} \\
		\subfloat[$\lambda = 0.05$]{\includegraphics[height = .3 \linewidth, width = .4 \linewidth]{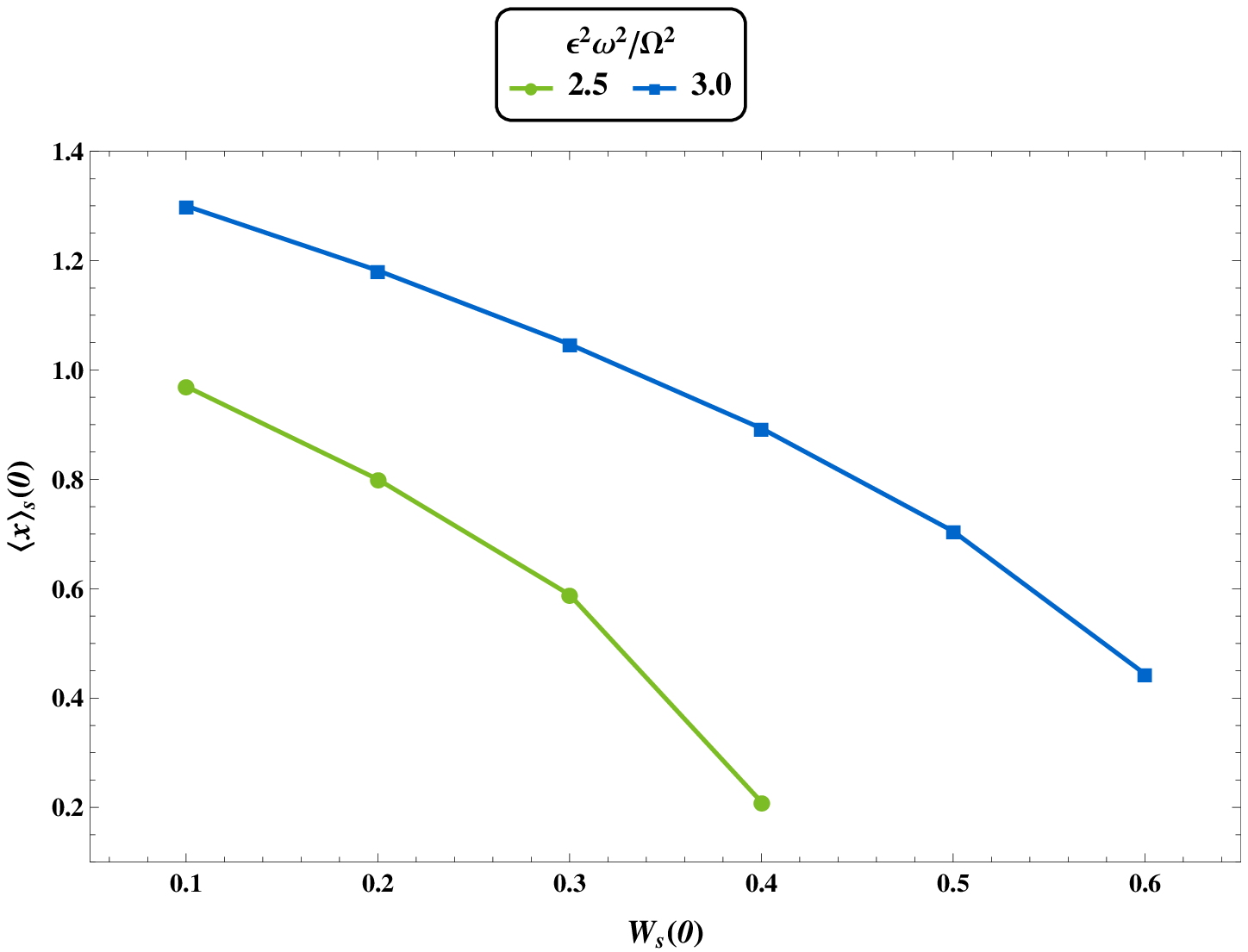} \label{fig:fig4c}} &
		\subfloat[$\lambda = 0.01$]{\includegraphics[height = .3 \linewidth, width = .4 \linewidth]{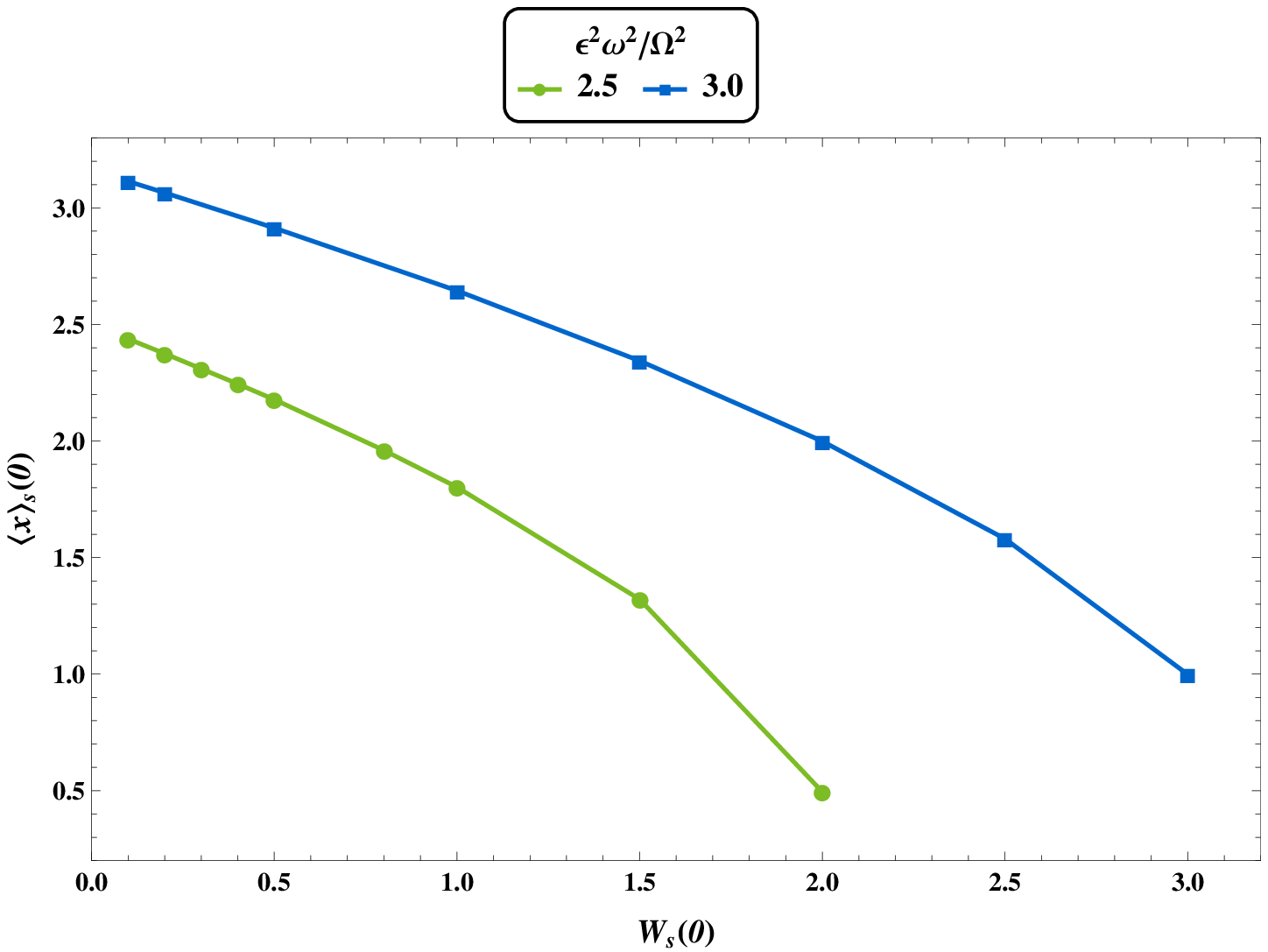} \label{fig:fig4d}}
	\end{tabular}
	\caption{Plots of maximum $\langle x \rangle_s(t = 0)$ versus maximum $W_s(t = 0)$ allowed for oscillations for $\omega^2 = 1$, $\frac{\epsilon^2 \omega^2}{\Omega^2} = \frac52, 3$ and various $\lambda$ for the partially coupled system (see discussions above). Notice that for a fixed initial mean square width, maximum permissible initial mean for bounded motion increases as $\lambda \rightarrow 0$. Also as $W_s(0) \rightarrow 0$, $\langle x \rangle_s(0)$ approaches the classical turning points.}
	\label{fig:fig4}
\end{figure}

Fig. \ref{fig:fig3} shows the time series of $\langle x \rangle_s$ for the same set of parameters as in Fig. \ref{fig:fig2}. Notice here the maximum initial $\langle x \rangle_s$ that the particle can have before escaping is less than for the uncoupled case discussed previously. We also observe that if we increase the initial mean square width of the wave packet keeping all other parameters fixed, the particle escapes the well much before $\langle x \rangle_s(t = 0)$ used for Fig. \ref{fig:fig3} is reached. If we keep on increasing initial $W_s$, we see that we need to decrease initial $\langle x \rangle_s$ for the particle to stay bounded. However if the initial $W_s$ is more than a critical value the particle escapes the well irrespective of how close to the center it is released. This is what is expected since increasing the initial $W_s$ amounts to increase in the quantum fluctuations. The phenomenon that is shown here very clearly is that the quantum fluctuations drive the particle to escape the well. Fig. \ref{fig:fig4} shows this scenario numerically, by plotting the maximum initial mean for which oscillation is possible against the initial mean square width for various $\lambda$ and fixed $\frac{\epsilon^2 \omega^2}{\Omega^2}$. We note that as $\lambda$ decreases the height of the potential as well as the turning points increases for both $\langle x \rangle_s$ and $W_s$. Thus higher the probability for the particle to be within the well for smaller $\lambda$. This is supported by the increasing maximum initial $\langle x \rangle_s$ as $\lambda$ decreases for a fixed initial $W_s$ and $\frac{\epsilon^2 \omega^2}{\Omega^2}$ as shown in Fig. \ref{fig:fig4}.

%%%%% PARTIALLY COUPLED SYSTEM %%%%%%%%%%%%
\onecolumngrid %sets to one column grid

\twocolumngrid %reverts to two column grid
%%%%%%%%%%%%% END PARTIALLY COUPLED SYSTEM %%%%%%%%%%%%%%%%

\phantom{} \phantom{} %invisible text to push text below

%%%%%%%%%%%%% START OF DISCUSSION FOR FULLY COUPLED SYSTEM %%%%%%%%%%%%%%%

\begin{figure}[h!]
	\includegraphics[width = \linewidth]{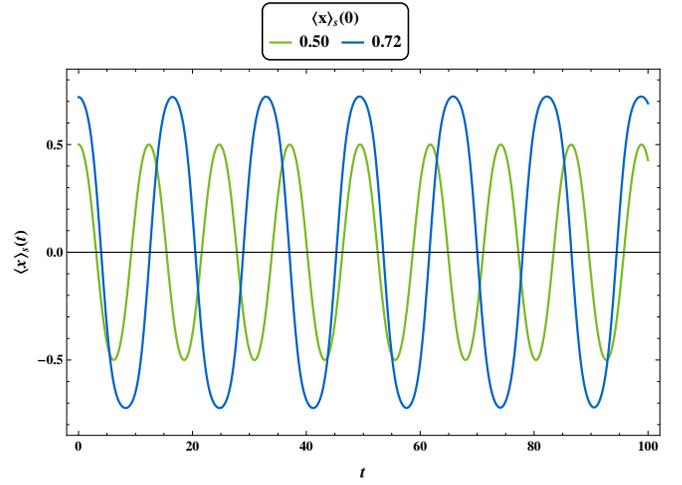}
	\caption{Numerical solution of the coupled dynamics of $\langle x \rangle_s$ with $\omega^2 = 1$, $\lambda = 0.1$, $\frac{\epsilon^2 \omega^2}{\Omega^2} = 3$ and $\langle x \rangle_s(0) = 0.5, 0.72$, $\langle \dot x \rangle_s(0) = 0$; with $W_s(0)= 0.1, \dot W_s(0) = 0$ and $\ddot W_s(0) = 0.01$.}
	\label{fig:fig5}
\end{figure}

\begin{figure}[h!]
	\begin{tabular}{cc}
		\subfloat[$\lambda = 0.5$]{\includegraphics[height = .3 \textwidth, width = .4 \textwidth]{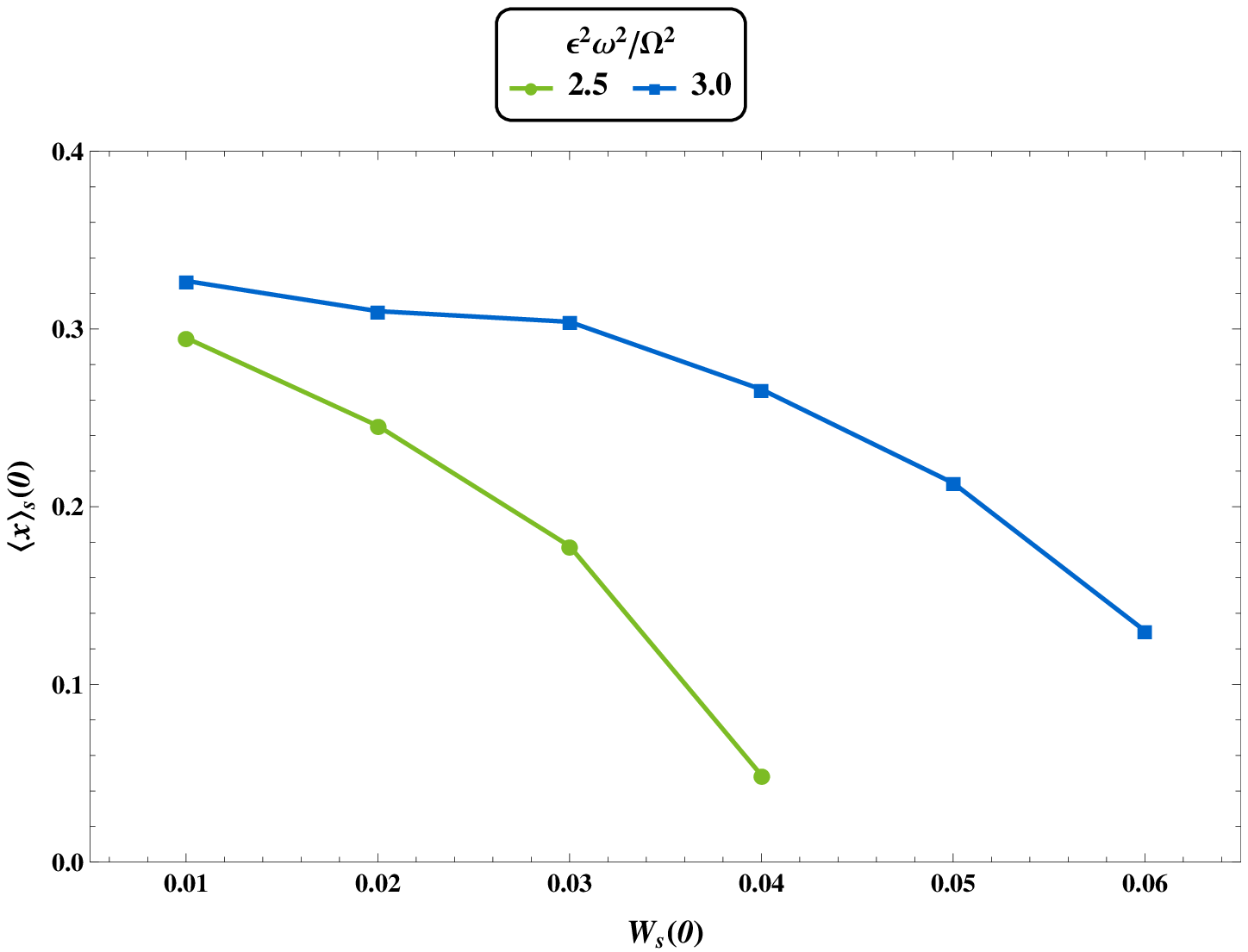} \label{fig:fig6a}} &
		\subfloat[$\lambda = 0.1$]{\includegraphics[height = .3 \textwidth, width = .4 \textwidth]{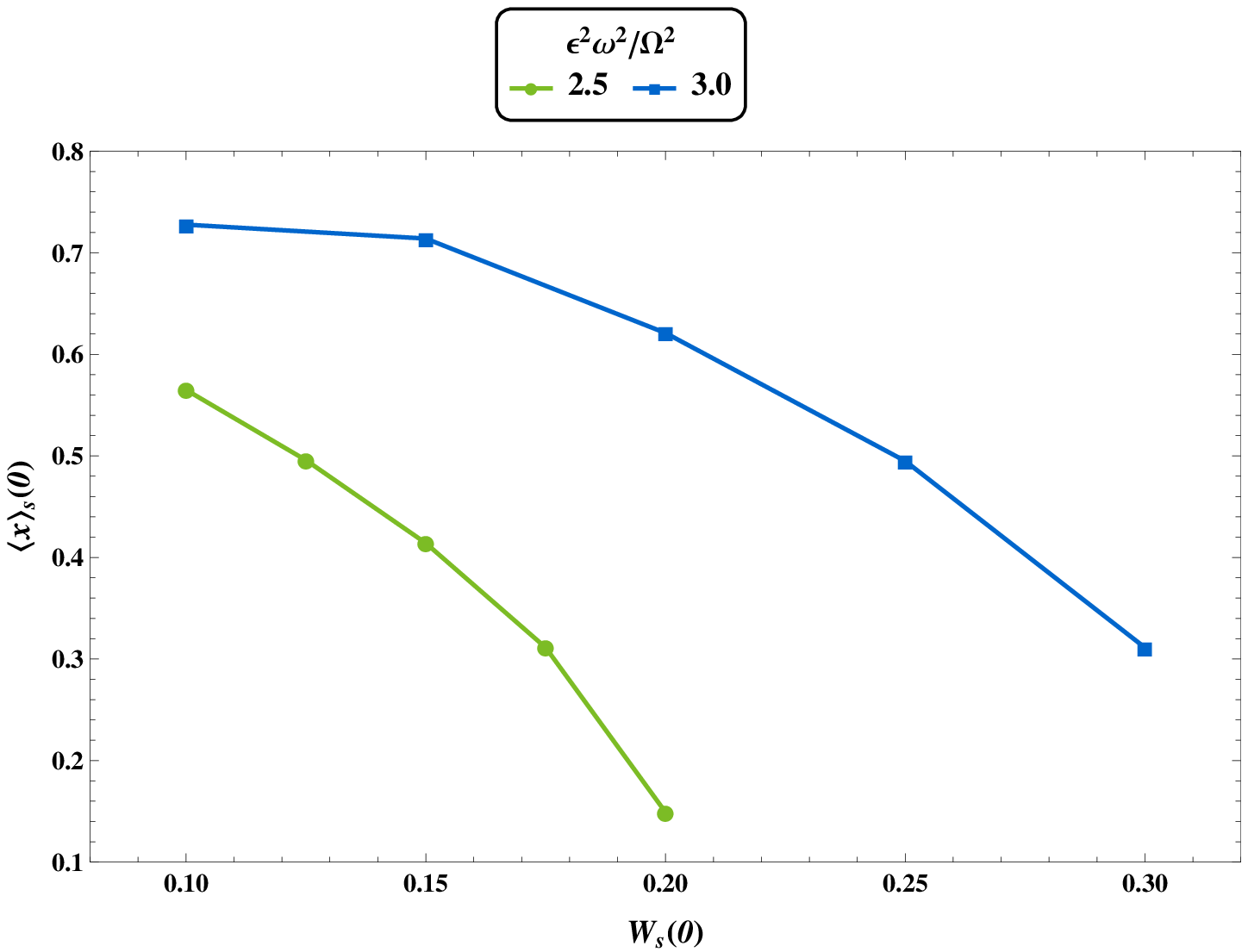} \label{fig:fig6b}} \\
		\subfloat[$\lambda = 0.05$]{\includegraphics[height = .3 \textwidth, width = .4 \textwidth]{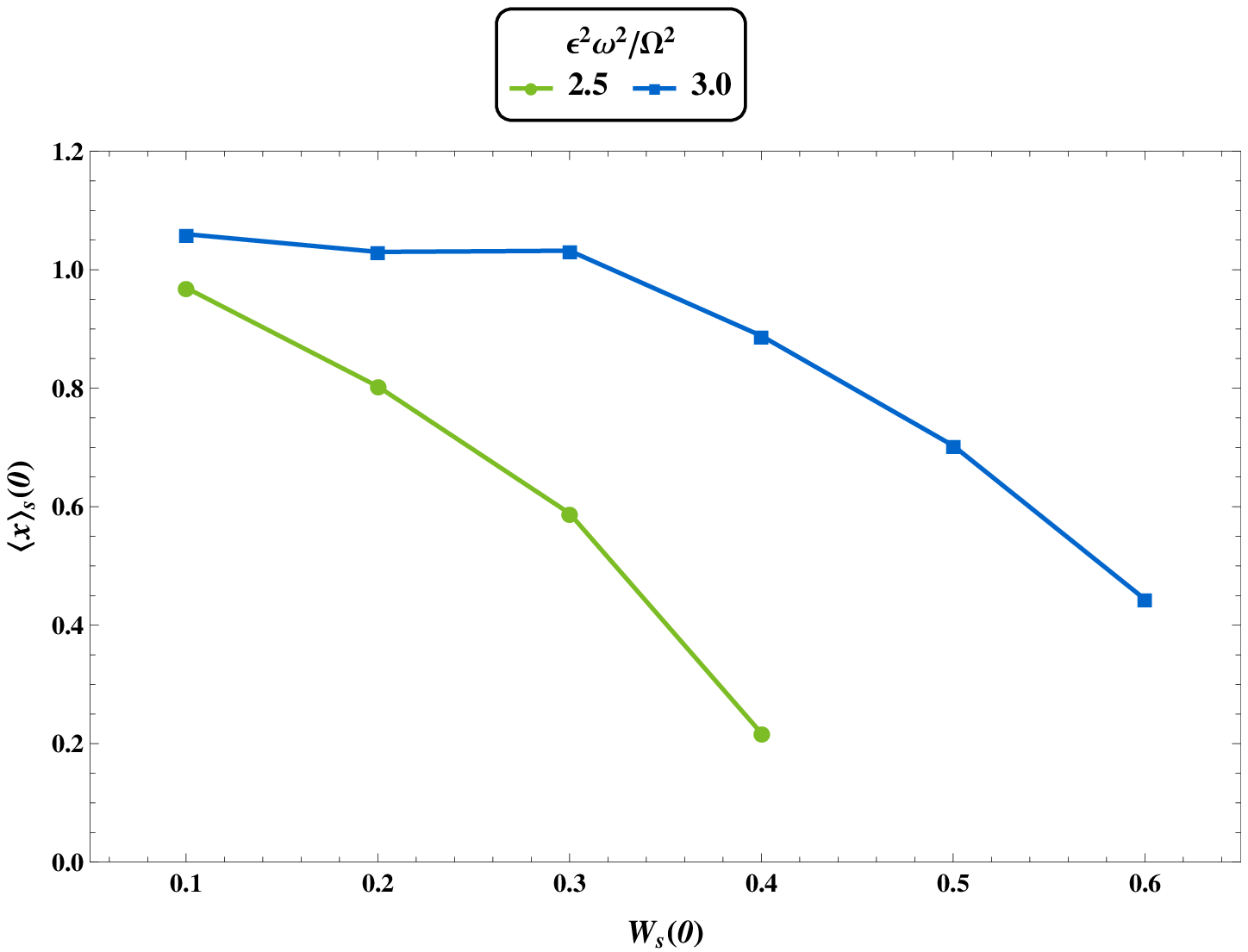} \label{fig:fig6c}} &
		\subfloat[$\lambda = 0.01$]{\includegraphics[height = .3 \textwidth, width = .4 \textwidth]{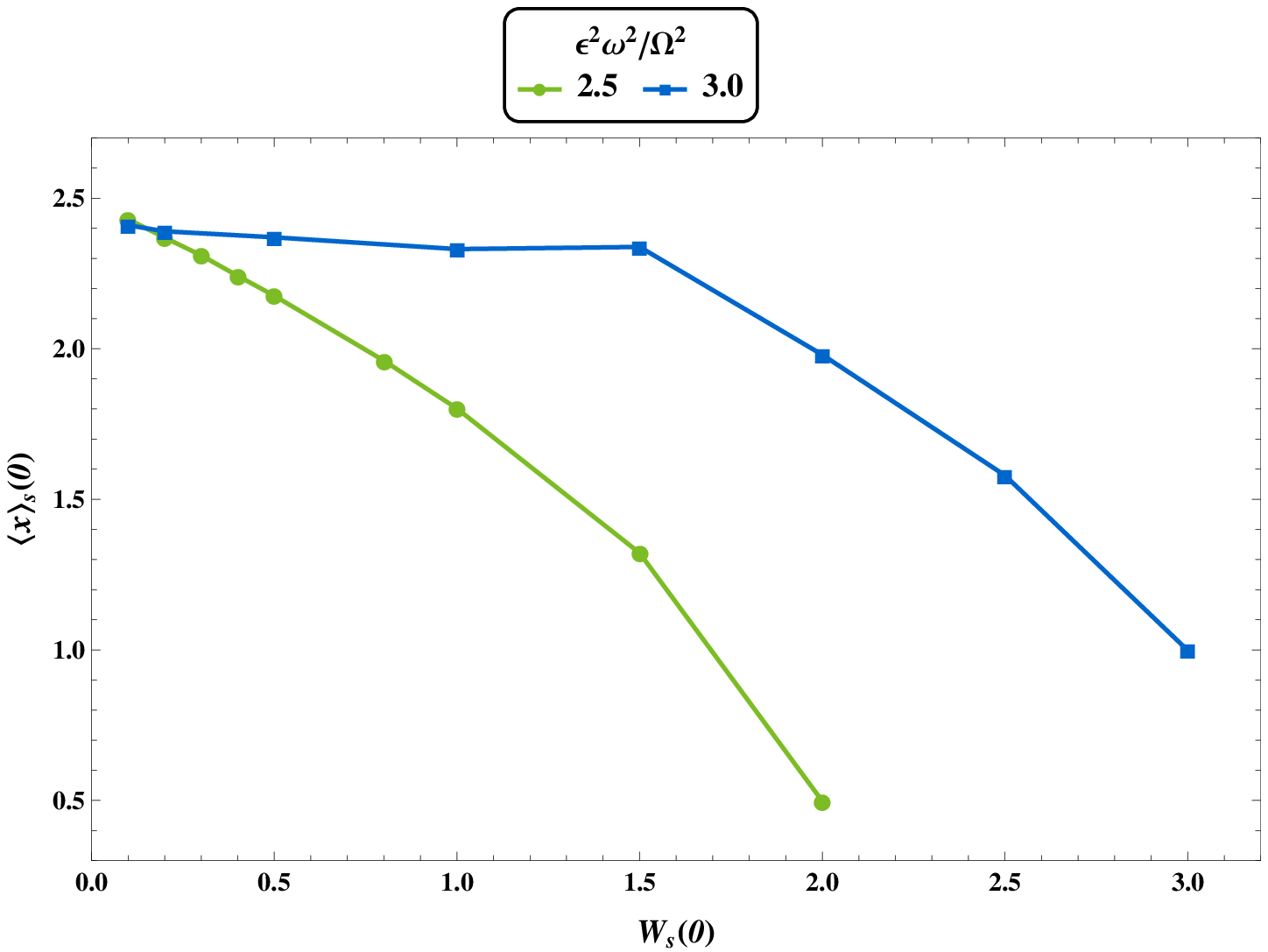} \label{fig:fig6d}}
	\end{tabular}
	\caption{Plots of maximum $\langle x \rangle_s(t = 0)$ versus maximum $W_s(t = 0)$ allowed for oscillations for $\omega^2 = 1$, $\frac{\epsilon^2 \omega^2}{\Omega^2} = \frac52, 3$ and various $\lambda$ for the coupled system as described by Eq. \eqref{eq:23} and Eq. \eqref{eq:24}. Notice that for a small $\lambda$ (Fig. \ref{fig:fig6c} and Fig. \ref{fig:fig6d}) $W_s(t = 0)$ has minimal effect on $\langle x \rangle_s(t = 0)$ below a critical value of $W_s(t = 0)$.}
	\label{fig:fig6}
\end{figure}

Finally we consider Eqs. \eqref{eq:23} and \eqref{eq:24} as is. Again we start with zero initial momentum and observe oscillations for the dynamics of $\langle x \rangle_s$. This is shown in Fig. \ref{fig:fig5}. We see a similar effect as in Fig. \ref{fig:fig3}. Plots for maximum initial mean versus initial mean square width possible for oscillations is shown in Fig. \ref{fig:fig6}. We also observe that there is minimal change to the maximum initial mean the particle can have beyond which it will escape when compared to the partially coupled system above (see Fig. \ref{fig:fig4}) as one increases the initial mean square width; thus confirming the small effect of the coupling in the dynamics of $W_s$. Also as $\lambda$ decreases the initial $W_s$ has a very small effect on initial $\langle x \rangle_s$ as shown by the flat nature of the graphs in Figs. \ref{fig:fig6c} and \ref{fig:fig6d}, for $\lambda = 0.05$ and $0.01$ respectively for $W_s(0)$ below some critical value.

%%%%%%%%%%%%%%% NUMERICAL PLOT OF FULLY COUPLED SYSTEM %%%%%%%%%%%%%%

%%%%% FULLY COUPLED SYSTEM %%%%%%%%%%%%
\onecolumngrid %sets to one column grid

\twocolumngrid %reverts to two column grid
%%%%%%%%%%%%% END FULLY COUPLED SYSTEM %%%%%%%%%%%%%%%%

\phantom{} \phantom{} %invisible text

At this point it is necessary to point out a shortcoming of our method of trying to understand the quantum dynamics in this situation. Undoubtedly, this is the simplest(but never before used) technique of capturing the quintessential feature of the quantum process, where the higher moments of the position and momentum operators play a very important role. This is what allows the tunneling to take place. However, as can be seen from Figs. \ref{fig:fig4c} and \ref{fig:fig4d} for small values of $\lambda$, it is very difficult to tunnel if the initial width of the wave packet is made very small. Tunneling occurs in this case only if the center of the wave packet is initially placed quite close to the location of the peak of the volcano. However quantum mechanics would require the particle to tunnel regardless of where the initial mean position of the particle is. It is natural to ask which part of our assumption is to blame. The Gaussian approximation for the kurtosis and the higher moments cannot be qualitatively wrong. It is the assumption that the skewness remains zero forever is what cannot be qualitatively right. As the mean position, $\langle x \rangle$, increases the skewness has to develop and will again tend to zero as the mean position becomes very large. Accordingly, we try the Ansatz where the skewness $S \equiv \left \langle \left (x - \langle x \rangle \right )^3 \right \rangle$ vanishes when $\langle x \rangle$ is very close to zero and is approximated by

\begin{equation} \label{eq:25}
	S = \gamma \langle x \rangle
\end{equation}

where $\gamma$ is an adjustable parameter. For small $\langle x \rangle$, small change in $\gamma$ can make the particle escape the well.  Now without going into the detailed calculation as we had done in Section \ref{quantum}, we see that in the last term of Eq. \eqref{eq:23}, $W_s$ will be changed to $W_s + \frac{S}{3 \langle x \rangle_s}$. Now if we simply take $S$ as defined above and also take $\langle x \rangle$ as $\langle x \rangle_s$, then the dynamics becomes

\begin{align} \label{eq:26}
\ddot{\langle x \rangle}_s &= \omega^2 \left( 1 - \frac{\epsilon^2 \omega^2}{2 \Omega^2} \right) \langle x \rangle_s - \lambda \left( 1 - \frac{2 \epsilon^2 \omega^2}{\Omega^2} \right) \langle x \rangle_s^3 \nonumber \\
& \quad - \lambda \left( 1 - \frac{2 \epsilon^2 \omega^2}{\Omega^2} \right) \big ( 3 \langle x \rangle_s W_s + \gamma \langle x \rangle_s \big )
\end{align}

For appropriate choice of $\gamma$ the particle can escape(via tunneling) for initial release close to the center of the well and small initial mean square width. We note that when $W_s$ is small, the particle is more \textit{classical}, and the skewness helps the particle to tunnel. Qualitatively according to Eq. \eqref{eq:12} the quantum mechanical phenomenon of tunneling is an effect prompted by the width of the wave packet and the skewness $S$ that it can develop in the course of time. We worked primarily under the assumption that $S$ remains zero at all times if it is initially so. When the width $W$ is made particularly small so that it is not sufficient by itself to cause tunneling we need to track the evolution of $S$ from zero to non-zero values to get the particle to escape the well.

\phantom{}

\onecolumngrid %sets to one column grid

\begin{figure}[b!]
	\begin{tabular}{cc}
		\subfloat[Classical system]{\includegraphics[height = .3 \linewidth, width = .4 \linewidth]{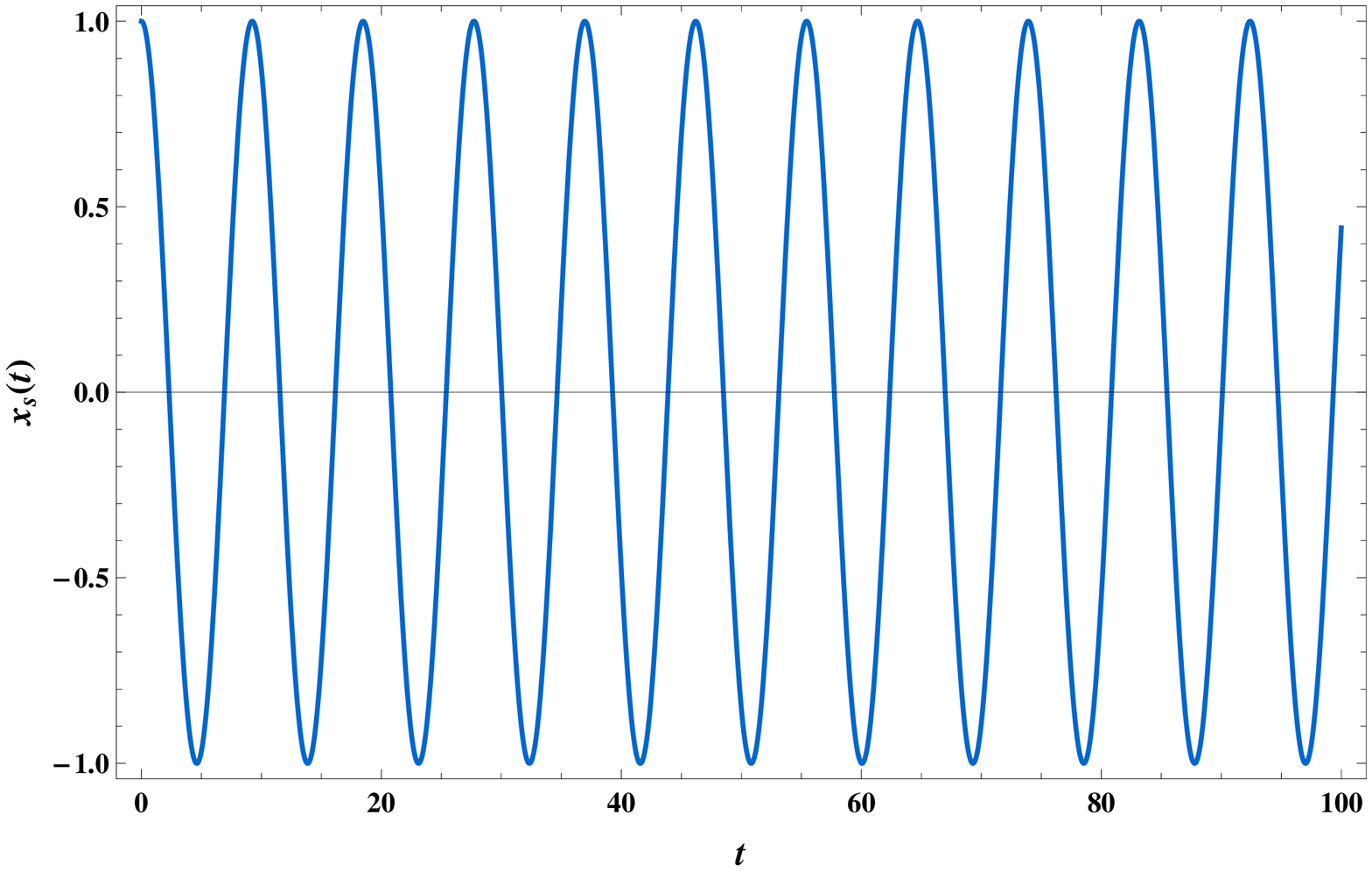} \label{fig:fig7a}} &
		\subfloat[Quantum system]{\includegraphics[height = .3 \linewidth, width = .4 \linewidth]{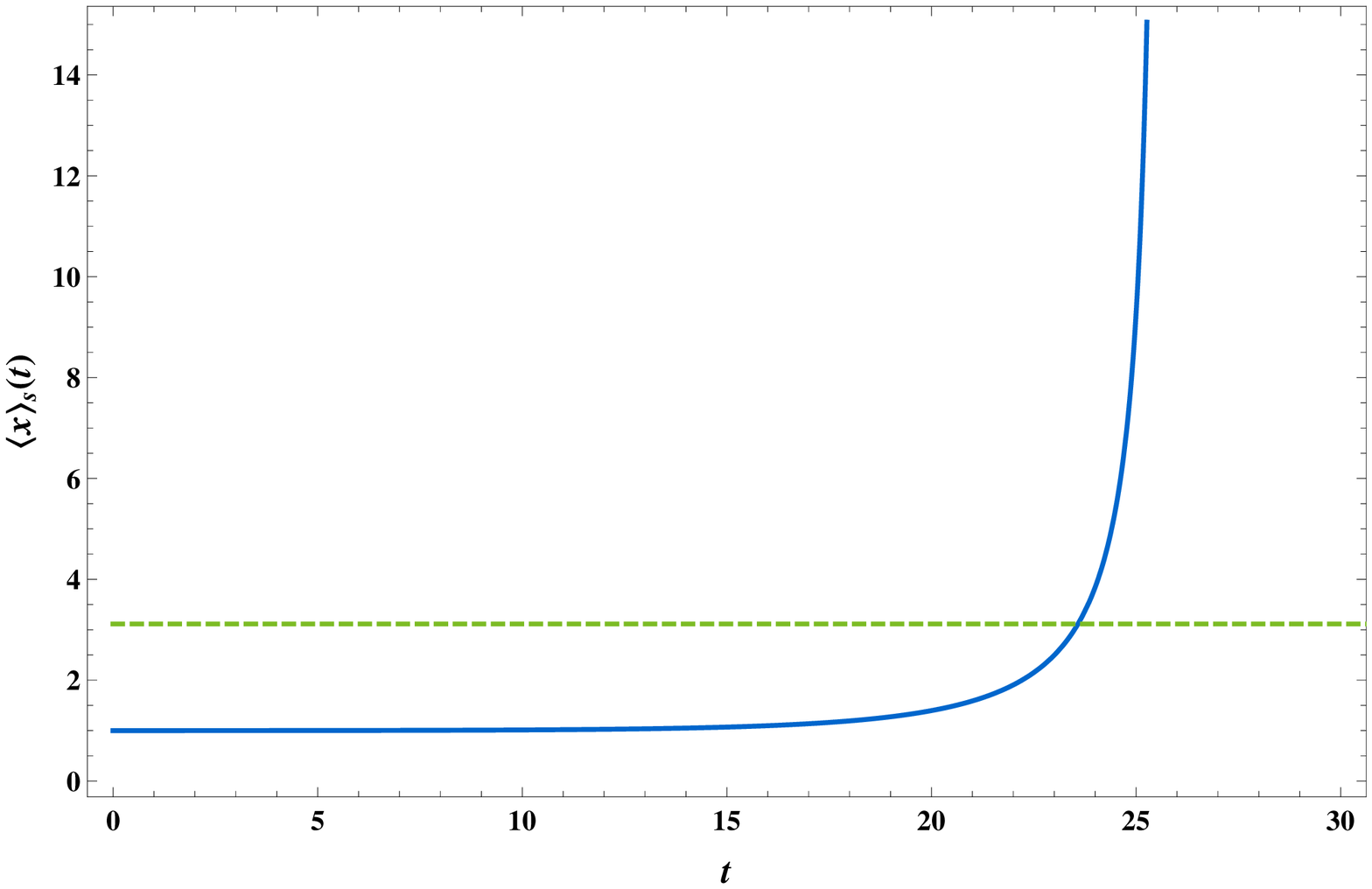} \label{fig:fig7b}} 
	\end{tabular}
	\caption{Numerical solution of (a) Eq. \eqref{eq:9} and (b) Eqs. \eqref{eq:23} and \eqref{eq:24} ($\langle x \rangle_s$ coupling ignored here) with $\omega^2 = 1$, $\lambda = 0.01$ and $\frac{\epsilon^2 \omega^2}{\Omega^2} = 3$. Both have the same initial conditions, with initial release from $x_s(0) = \langle x \rangle_s(0) = 1$ and zero initial momentum. Additionally for (b) $W_s(0)= 3.0, \dot W_s(0) = 0$ and $\ddot W_s(0) = 0.01$. The mean square width here kicks the particle out of the well though its classically bounded. The green dashed line shows the classical turning point $3.16$, for the given parameters.}
	\label{fig:fig7}
	\medskip	
\end{figure}

\twocolumngrid %reverts to two column grid

\begin{figure}[b!]
	\begin{tabular}{cc}
		\subfloat[]{\includegraphics[height = .3 \linewidth, width = .4 \linewidth]{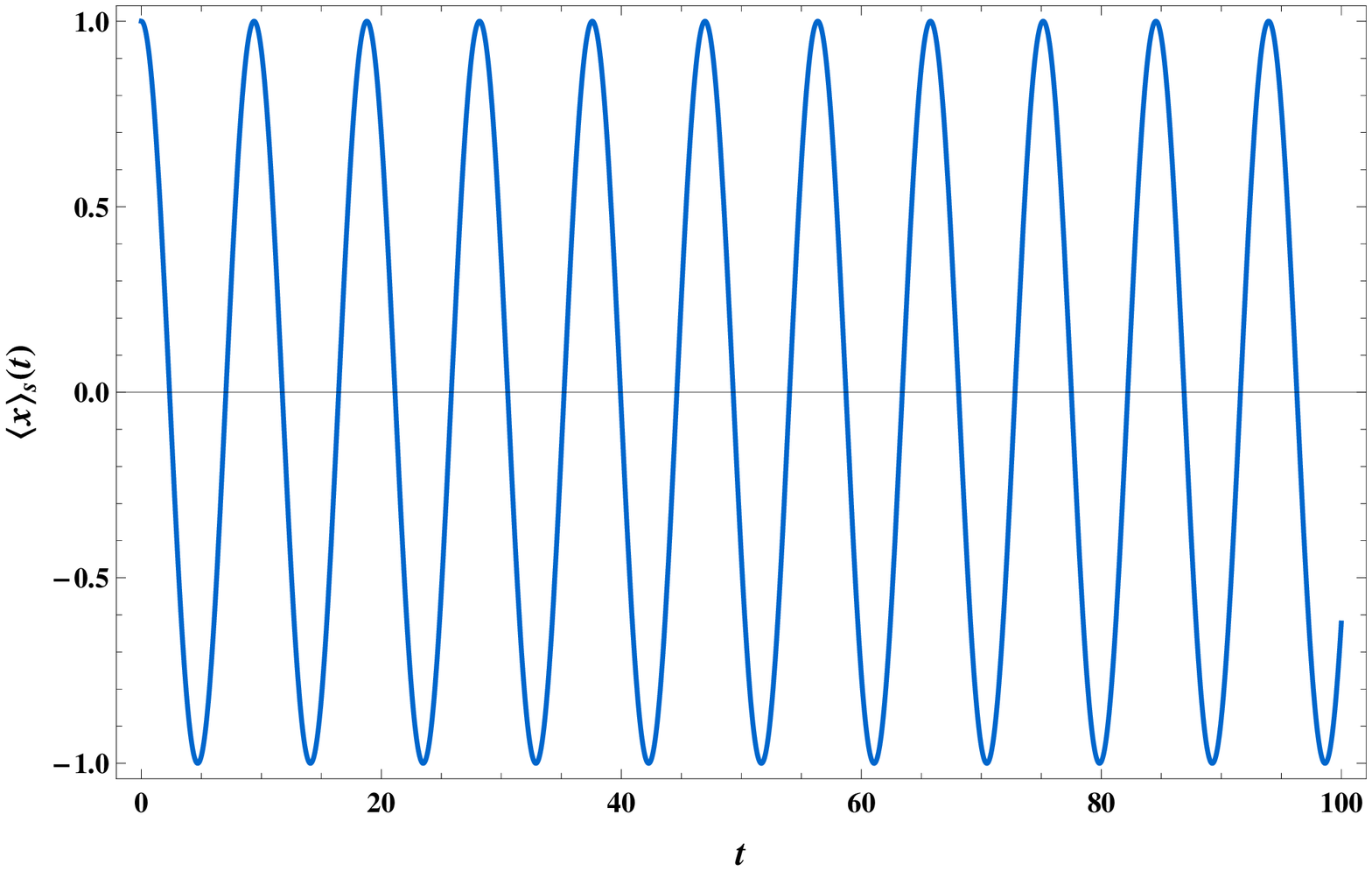} \label{fig:fig8a}} &
		\subfloat[]{\includegraphics[height = .3 \linewidth, width = .4 \linewidth]{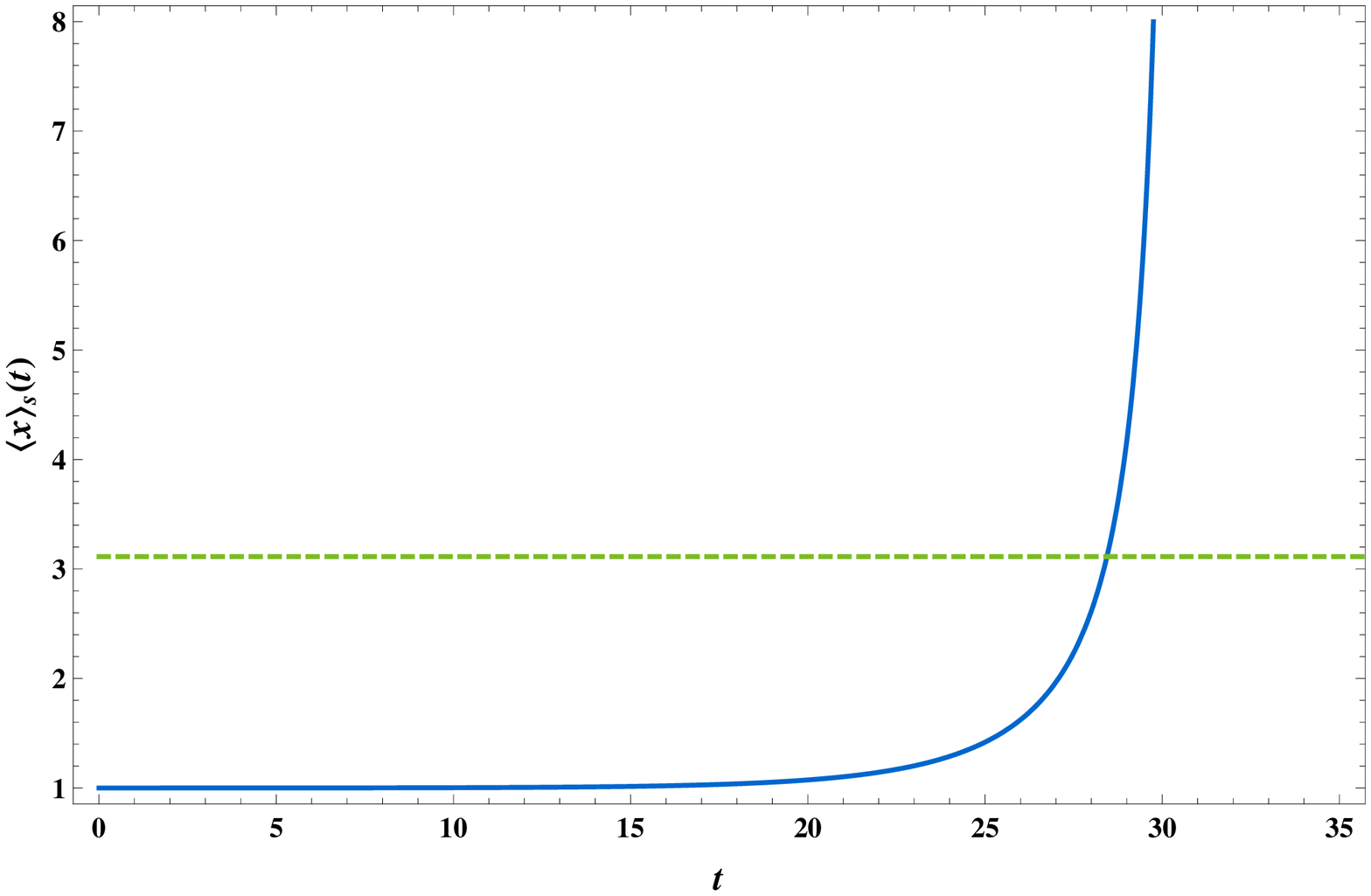} \label{fig:fig8b}} 
	\end{tabular}
	\caption{Numerical solution of (a) Eq. \eqref{eq:23} and (b) Eq. \eqref{eq:26} and Eq. \eqref{eq:24} ($\langle x \rangle_s$ coupling ignored here) with $\omega^2 = 1$, $\lambda = 0.01$ and $\frac{\epsilon^2 \omega^2}{\Omega^2} = 3$. Both have the same initial conditions, with initial release from $\langle x \rangle_s(0) = 1$ and zero initial momentum; additionally $W_s(0)= 0.1, \dot W_s(0) = 0$ and $\ddot W_s(0) = 0.01$. Here in (b) $\gamma = 8.699$, which assists the particle to tunnel even though in (a) its bounded. We see that skewness indeed helps the particle to escape. The green dashed line shows the classical turning point $3.16$, for the given parameters.}
	\label{fig:fig8} 
	\bigskip % give some vertical space
\end{figure}

We see in Fig. \ref{fig:fig7} that for the same initial conditions, the quantum fluctuations (which is provided here by $W_s$) makes the motion unbounded though classically it is stable. Now we add skewness into the picture  and anticipate that proper choice of $\gamma$ in Eq. \eqref{eq:26} will help the particle escape if we initially release the particle close to the center and also for small initial mean square width. Fig. \ref{fig:fig8} shows that for proper choice of the parameters of skewness $S$ (see Eq. \eqref{eq:25}), the particle escapes the well.

\phantom{} \phantom{} \phantom{} %gives some blanck spaces 

\onecolumngrid %sets to one column grid

\twocolumngrid %reverts to two column grid

%%%%%%%%%%%%%%%% CONCLUSION %%%%%%%%%%%%%%%%
\section{Conclusion} \label{conclusion}

Summarizing, we have shown that an effective volcano potential can be generated from a periodically forced double well potential when the amplitude and frequency of forcing satisfy certain constraints. For a classical particle this implies that the particle can escape from the confining potential if its initial energy is greater than some critical value. For a quantum particle, we have seen that  this implies escape virtually for any initial condition because of the tunneling effect. In our way of doing this dynamics we have captured the escape by looking at the evolution of the mean position and the width of an initial wave packet. The coupling between the mean position and the width allows the quantum particle to escape even when the classical is bound. This is in conformity with the fact that the quantum states in the volcano potential are resonance states.

%%%%%%%%%%%%%%% BIBLIOGRAPHY %%%%%%%%%%%%%%%5

\end{document}